\documentclass[a4paper]{article}
\usepackage[pdftex]{graphicx} 
\usepackage[english]{babel}
\usepackage[bf,small,nooneline]{caption}
\usepackage{multirow}
\usepackage[round]{natbib}

\newcommand{\dfrac}[2]{\frac{\displaystyle#1}{\displaystyle#2}}
\newcommand{\pder}[2]{\frac{\displaystyle\partial#1}{\displaystyle\partial#2}}
\newcommand{\OOp}{\mathrm{O}_2^+}
\newcommand{\COOp}{\mathrm{CO}_2^+}
\newcommand{\hv}{h\nu}

\title{Details of a Hybrid Model for the Interaction between the Solar Wind and Planets Implemented in FLASH} 
\begin{document}
\date{September 25, 2015}
\author{M.\ Holmstr{\"o}m\thanks{Swedish Institute of Space Physics, PO~Box~812, SE-98128~Kiruna, Sweden. (\texttt{matsh@irf.se})}, and X.-D.\ Wang$^*$}
\maketitle

\begin{abstract}
A hybrid plasma solver treats ions as particles and electrons as a fluid.  
We have implemented a parallel hybrid solver in the FLASH open source software framework.  The solver has been applied to studies of the interaction between the solar wind and planets. 
Here we discuss the implementation of different model features, such as permanent magnetic fields, ionospheric chemistry, and exospheres.  
Mars is used as an example. 
\end{abstract}

\section{Introduction to the hybrid model} 
In the hybrid approximation, ions are treated as particles, 
and electrons as a massless fluid. 
In what follows we use SI units. 
The trajectory of an ion, $\mathbf{r}(t)$ and $\mathbf{v}(t)$, 
with charge $q$ and mass $m$, is computed from the Lorentz force, 
\[
  \dfrac{d\mathbf{r}}{dt} = \mathbf{v}, \quad
  \dfrac{d\mathbf{v}}{dt} = \dfrac{q}{m} \left( 
    \mathbf{E}+\mathbf{v}\times\mathbf{B} \right), 
\]
where $\mathbf{E}=\mathbf{E}(\mathbf{r},t)$ is the electric field, 
and $\mathbf{B}=\mathbf{B}(\mathbf{r},t)$ is the magnetic field.  
From now on we do not write out the dependence on $\mathbf{r}$ and $t$. 
The electric field is given by 
\begin{equation}
  \mathbf{E} = \dfrac{1}{\rho_I} \left( -\mathbf{J}_I\times\mathbf{B} 
  +\mu_0^{-1}\left(\nabla\times\mathbf{B}\right) \times \mathbf{B} 
   - \nabla p_e \right) + \frac{\eta}{\mu_0} \nabla\times\mathbf{B}, 
\label{eq:E}
\end{equation}
where $\rho_I$ is the ion charge density, 
$\mathbf{J}_I$ is the ion current density, 
$p_e$ is the electron pressure, 
$\eta$ is the resistivity, and
$\mu_0=4\pi\cdot10^{-7}$ is the magnetic constant. Then 
Faraday's law is used to advance the magnetic field in time, 
        \[
          \pder{\mathbf{B}}{t} = -\nabla\times\mathbf{E}. 
        \]
Note that the unknowns are the position and velocity of the ions, and 
the magnetic field on a grid, \emph{not} the electric field, since it 
can always be computed from~(\ref{eq:E}). As is common for hybrid models, 
we use the Darwin approximation~\citep{Schmitz06}, 
i.e.\ we neglect the displacement current 
in Ampere's law, and the current then is 
$\mathbf{J}=\mu_0^{-1}\nabla\times\mathbf{B}$. 
This removes waves that propagate with the speed of light, $c$. 
For the problems we study the velocities are small compared to $c$  
and the transport of energy by electromagnetic radiation
can be neglected.  This makes the approximation reasonable for our purposes. 
If the Darwin approximation is not used we would have the 
severe restriction on the time step of $\Delta t < \Delta x/c$. 

We have implemented a parallel hybrid solver in the FLASH open source software framework~\citep{Fryxell00}.  
Further details on the hybrid model used here, and the discretization, 
can be found in~\citet{Enumath09,Astronum10,Astronum12}. 
The solver has been applied to studies of the interaction between the solar wind and planets, e.g., the Moon~\citep{EPS}. 

In the following sections we present some details of this hybrid model. 
How we handle permanent magnetic fields, how ionospheric chemistry for 
Mars can be included in the model, and some analytical expressions 
for Mars' exosphere. 

\section{Permanent magnetic fields} 
In many applications there is a background magnetic field that does not change over time.  An example can be the internal dipole field of a planet, or the solar wind magnetic field.  For simplicity and accuracy over long time runs we can split the field into a time varying part and a constant in time part~\citep{Tanaka}. 
We can then solve for $\mathbf{B}(\mathbf{r},t)=\mathbf{B}_0(\mathbf{r})+\Delta\mathbf{B}(\mathbf{r},t)$. Note that $\Delta\mathbf{B}$ does not need to be small. To ensure that $\nabla\cdot\mathbf{B}=0$ we can define $\mathbf{B}_0(\mathbf{r})=-\nabla\psi(\mathbf{r})$, then we also have that $\nabla\times\mathbf{B}_0=0$. 
Inserting this into the hybrid equations, we find that $\mathbf{B}_0$ then only appears in the Lorentz force and in the first term of the electric field expression.  All other terms in the electric field expression involving $\mathbf{B}_0$ are zero since $\mathbf{B}_0$ is rotation free. 

To implement this in the hybrid solver it is important that $\nabla\cdot\mathbf{B}_0=0$ also in the discrete sense.  This can be ensured by computing the background field from the potential, $\mathbf{B}_0(\mathbf{r})=-\nabla\psi(\mathbf{r})$ using the same finite difference approximation of the gradient that we use for the divergence.  In our case we use the same centered finite difference stencil.  This ensures that the discrete $\nabla\cdot\mathbf{B}_0=0$ down to round-off errors.  Note that $\mathbf{B}_0$ need only be computed once, at the start of the simulation, and stored. 

As an example, the potential for a dipole is \[
\psi(\mathbf{r}) = \frac{\mu_0}{4\pi} \frac{\mathbf{m}\cdot\mathbf{r}}{r^3}  =   \frac{\mathbf{m}\cdot\mathbf{r}}{r^3}  \cdot 10^{-7}    \mbox{ [T m].} 
\] 
Then
\[
\mathbf{B}_d = -\nabla\psi(\mathbf{r})  =  \frac{\mu_0}{4\pi r^3} \left(  3  \frac{ \mathbf{r} \left(  \mathbf{m}\cdot\mathbf{r}  \right) }{r^2} -  \mathbf{m} \right)  = \frac{10^{-7}}{r^3} \left(  3  \frac{ \mathbf{r} \left(  \mathbf{m}\cdot\mathbf{r}  \right) }{r^2} -  \mathbf{m} \right). 
\]
Here $\mathbf{m}$ is the dipole moment [T m$^3$].  The background field from a dipole located at $\mathbf{r}_c$ then is $\mathbf{B}_0(\mathbf{r})=\mathbf{B}_d( \mathbf{r} - \mathbf{r}_c )$. 

For a conducting sphere of radius $R$ in a uniform field $\mathbf{B}_u$, currents in the sphere will expel the external field such that the component normal to the sphere surface is zero.  The resulting total field outside the sphere can be seen as the sum of the external field and a dipole at the center of the sphere directed opposite to the external field.   The magnitude of the required dipole moment is $m = 10^7 R^3 B_u / 2$. 

\section{Ionospheric chemistry for Mars}
\begin{figure}
\begin{center}
  \includegraphics[width=0.8\columnwidth]{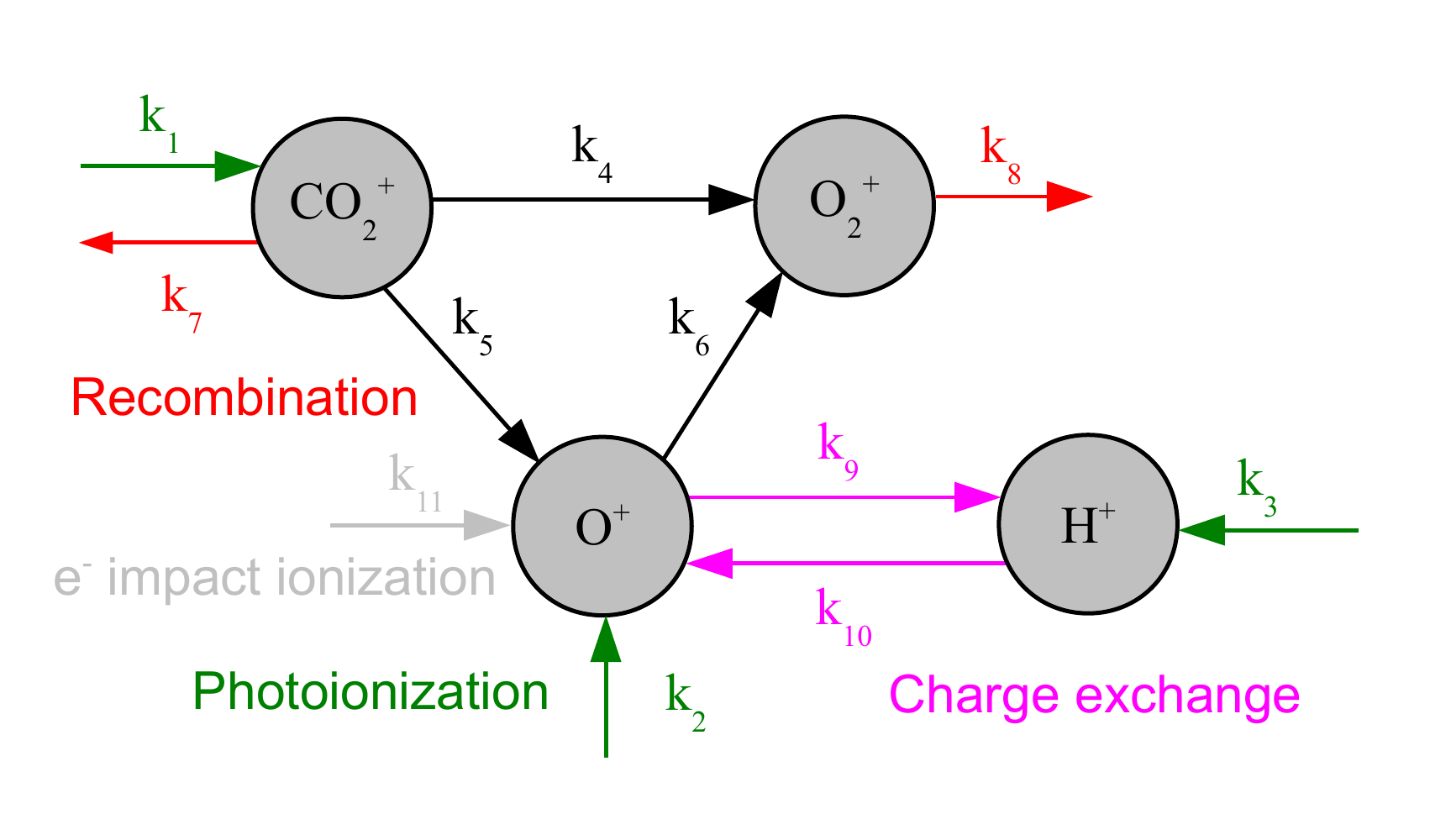}
\end{center}
\caption{  An illustration of the ionospheric chemical reactions considered 
here for the four ion species, following~\citet{Ma} and~\citet{Brecht}. 
Sources of each ion species are shown by incoming arrows. 
Outgoing arrows show losses and connecting arrows show how one 
species is transformed into another.  
Red color for photoionization and green for recombination with an electron. 
The rate for each reaction is $k_i=k_i(\mathbf{r}), i=1,\ldots,11$, 
and values are given in Table~\ref{table:reactions}. 
Some rates depend on photon flux, neutral temperatures, 
and electron temperature.
} \label{fig:ions} 
\end{figure}

For Mars ionosphere one can consider only the major ion species 
CO$_2^+$, O$^+$, and O$_2^+$. 
A table of the most relevant chemical reactions involving these species was 
presented by~\citet{Ma}.  A similar table is provided by~\citet{Brecht}, 
with the addition of oxygen electron impact ionization, 
e+O~$\rightarrow$~O$^+$+e+e.  
The different reactions are presented in Figure~\ref{fig:ions} 
and in Table~\ref{table:reactions}. 
We also consider the hydrogen exosphere, but H only occur for photoionization and charge exchange. 

We can note that all reactions depend only on one ion species, 
so we can adopt a Monte Carlo approach where for each meta-ion we can 
compute the probability that it undergo a reaction and then randomly 
convert it to another ion. 
Recombination reactions also depend on the electron density 
(total charge density) so we have to get that value from the grid. 

The initial ion densities are important. 
We disregard charge exchange for initial distributions. 

Here square brackets denote number density of a species, e.g., 
$[A]$~m$^{-3}$ for a species $A$. 
We consider the density of neutrals as constant in radial background profiles 
\[
   [\textrm{H}](r), [\textrm{O}](r), [\textrm{CO}_2](r)
\]
where the density only depends on distance, $r$, from the center of Mars. 
So the neutral atmosphere is an unchanging source and sink. 
This is an acceptable approximation at Mars to large distances from the 
planet since the neutral densities are so much larger than the ion densities. 

The unknowns here are the ion number densities as a function of position 
and time 
\[
   [\textrm{H}^+](\mathbf{r},t), [\textrm{O}^+](\mathbf{r},t), 
   [\textrm{O}_2^+](\mathbf{r},t), [\textrm{CO}_2^+](\mathbf{r},t). 
\]
We will also need the electron number density
\[
  [e](\mathbf{r},t) = [\textrm{H}^+] + [\textrm{O}^+] + 
                      [\textrm{O}_2^+] + [\textrm{CO}_2^+]. 
\]
Using the reactions in~\citet{Ma} we can then write the 
evolution of ion number densities at each position $\mathbf{r}$ 
(no transport is considered) as 
\begin{eqnarray*}
  \dfrac{d}{dt}[\textrm{H}^+] & = & k_3 [\textrm{H}] 
     + k_9 [\textrm{H}] [\textrm{O}^+] - k_{10} [\textrm{O}] [\textrm{H}^+] \\ 
  \dfrac{d}{dt}[\textrm{O}^+] & = & k_2 [\textrm{O}] 
     + k_5 [\textrm{O}] [\textrm{CO}_2^+] - k_6 [\textrm{CO}_2] [\textrm{O}^+] 
     - k_9 [\textrm{H}] [\textrm{O}^+] + k_{10} [\textrm{O}] [\textrm{H}^+] 
     + k_{11} [\textrm{O}] [\textrm{e}] \\
  \dfrac{d}{dt}[\textrm{CO}_2^+] & = & k_1 [\textrm{CO}_2] 
     - (k_4+k_5) [\textrm{O}] [\textrm{CO}_2^+] 
     - k_7 [\textrm{CO}_2^+] [\textrm{e}] \\
  \dfrac{d}{dt}[\textrm{O}_2^+] & = & k_4 [\textrm{O}] [\textrm{CO}_2^+] 
     + k_6 [\textrm{CO}_2] [\textrm{O}^+] - k_8 [\textrm{O}_2^+] [\textrm{e}]
\end{eqnarray*}
Some of the constants $k_i$ also depend on ion and electron temperature, 
$T_i$ and $T_e$. 
% Should depend on velocity of ion, not $T_e$, in MC approach. 
We list the coefficients in Table~\ref{table:reactions}.  All coefficients except for $k_{11}$ come from~\citet{Krasnopolsky}. 
The coefficient $k_{11}$ is re-cited from~\citet{Roberson}. 

\begin{table} 
\caption{Table of ion species, the reactions considered, the rate for each reaction, and the symbol used for the rate.}
\label{table:reactions}
\begin{center}
\begin{tabular}{llll}
\hline
 Species &Reaction& $^{a}$Rate  & Symbol \\
\hline
\multirow{6}{*}{O$^+$} & O$+\hv\rightarrow $O$^+ + e$ & $^{b}$8.89E-8,\, 2.73E-7  s$^{-1}$& $k_2$ \\
 & O$+\COOp\rightarrow $O$^++$CO$_2$& 9.6E-11 &$k_5$ \\
 & O+H$^+\rightarrow $O$^++$H & $^{c}$7.3E-10$(T/300)^{0.23}e^{-226/T}$ &$k_{10}$ \\
 &O$+e\rightarrow $O$^++e+e$& $^{ef}$1.29E-11$T_e^{0.7}e^{-1.58E5/T_e}$ &$k_{11}$ \\
 &O$^++$CO$_2\rightarrow \OOp+$CO& 9.4E-10 &$k_{6}$ \\
 &O$^++$H$\rightarrow $H$^++$O& $^{c}$5.7E-10$(T/300)^{0.36}$ &$k_{9}$ \\
\hline
\multirow{3}{*}{H$^+$} & H$+\hv\rightarrow $H$^++e$& $^{b}$5.58E-8,\, 8.59E-8 s$^{-1}$ &$k_3$ \\
&O$^++$H$\rightarrow $H$^++$O&$^{c}$5.7E-10$(T/300)^{0.36}$ &$k_9$ \\
&H$^++$O$\rightarrow $O$^++$H&$^{c}$7.3E-10$(T/300)^{0.23}e^{-226/T}$ &$k_{10}$ \\
\hline
\multirow{3}{*}{$\OOp$} &O$+\COOp\rightarrow \OOp+$CO& 1.64E-10 &$k_4$ \\
&CO$_2+$O$^+\rightarrow \OOp+$CO& 9.4E-10 &$k_6$ \\
&$\OOp+e\rightarrow$ O+O&$^{d}$2E-7$(300/T_e)^{0.5}$ &$k_8$ \\
\hline
\multirow{4}{*}{$\COOp$}& CO$_2+\hv\rightarrow \COOp+e$&$^{b}$2.47E-7,\, 7.3E-7 s$^{-1}$ &$k_1$ \\
&$\COOp+$O$\rightarrow $O$^++$CO$_2$& 9.6E-11 &$k_5$ \\
&$\COOp+$O$\rightarrow \OOp+$CO& 1.64E-10 &$k_4$ \\
&$\COOp+e\rightarrow$ CO+O& $^{d}$3.8E-7$(300/T_e)^{0.5}$ &$k_7$ \\
\hline
\end{tabular}

$^a${in cm$^3$s$^{-1}$, unless specified.}
$^b${solar min., solar max.}
$^c${$T$ the neutral temperature in~K.}
$^d${$T_e$ the electron temperature in K.}
\end{center}
\end{table}

\begin{table}
\caption{Table of reaction times when considering individual ions in a Monte Carlo approach.  The number density of species $M$ is denoted $[M]$ and is given in cm$^{-3}$}
\label{table:times}
\begin{center}
\begin{tabular}{lll}
\hline
Reaction& Effect &  Time \\
\hline
O$^+$+CO$_2\rightarrow\OOp$+CO & O$^+$ converted to O$_2^+$  & $(k_6\cdot[\mathrm{CO}_2])^{-1}$ \\
O$^++$H$\rightarrow $H$^++$O & O$^+$ converted to H$^+$ & $(k_9\cdot[\mathrm{H}])^{-1}$ \\
H$^++$O$\rightarrow $O$^++$H & H$^+$ converted to O$^+$ & $(k_{10}\cdot[\mathrm{O}])^{-1}$ \\
$\OOp+e\rightarrow $O+O & Loss of $\OOp$ & $(k_{8}\cdot[e])^{-1}$ \\
$\COOp+$O$\rightarrow $O$^++$CO$_2$ & $\COOp$ converted to O$^+$ & $(k_{5}\cdot[\mathrm{O}])^{-1}$ \\
$\COOp+$O$\rightarrow \OOp+$CO & $\COOp$ converted to $\OOp$ & $(k_{4}\cdot[\mathrm{O}])^{-1}$ \\
$\COOp+e\rightarrow $CO+O & Loss of $\COOp$ & $(k_{7}\cdot[e])^{-1}$ \\
\hline
\end{tabular}
\end{center}
\end{table}

The chemistry can be implemented using a Monte Carlo approach. 
Each existing ion in the simulation has an average time for each 
reaction.  If we denote such a time $\tau$, after each time step, 
for each meta-particle, we draw a random time from an exponential 
distribution with mean $\tau$.  The event occur if this time is 
smaller than the time step. 
For photoionization these times are listed in Table~\ref{table:reactions}. 
The times for the other reactions are listed in Table~\ref{table:times}. 

At the beginning of the simulation we could start with no ionospheric 
ions and let them evolve over time using the Monte Carlo approach. 
It could however take too long time to reach a steady state. 
We could instead initialize the ionospheric densities in our simulation model 
to values close to the steady state. 
Exact values are not needed since ion transport will anyway change the 
state over time. 
Then we need to compute the steady state number densities of ions.  
This can be done by integration in time 
until a steady state is reached~\citet{Brecht}. 
Alternatively, setting the time derivatives to zero above gives us 
a system of non-linear equations, the solution of which is the  
steady state solution.  The equations are non-linear due to the 
presence of the electron density in the equations. 
It is not possible to compute a simple 
analytical solution to the above equations. %  See syms.txt 
It is however possible to find the solution numerically. 
This is similar to using an implicit in time solver to 
compute the steady state solution.

\section{Mars' exosphere}
There are very few measurements of the hydrogen exosphere of Mars, all of which are UV remote sensing experiments \citep{anderson1971, anderson1974, anderson1976, chaufray2008, feldman2011}. Among these experiments, most observations were been done with the Lyman-$\alpha$ line except for the Rosetta/ALICE observation \citep{feldman2011}, that employed the Lyman-$\beta$ line as well.  Although the Lyman-$\alpha$ emission is the strongest ($\approx 250$ times stronger than that of Lyman-$\beta$), it is not optically thin below 10000 km altitude \citep{anderson1976, feldman2011}, meaning that one has to model the radiative transfer processes in the hydrogen exosphere to resolve the density profile, and the result is usually not unique.  The density profiles resolved from Lyman-$\alpha$ observations consist of two components: a cold or thermal population with a temperature of $\approx 200$~K at the exobase, consistent with Chamberlain's theory of exosphere formation, and a hot or supra-/non thermal population with a temperature of more than 500~K at the exobase, the production mechanism of which is yet unclear. The Lyman-$\beta$ line, on the other hand, is optically thin except for the lowest altitudes at the planetary limb. This allows to interpret the emission intensity by the simple integration along the line of sight therefore the density profile is better constrained. The altitude profiles derived by the Lyman-$\beta$ observation does not include the hot population. In our current study we take the one from the Lyman-$\beta$ line observation by Rosetta/ALICE due to its more straightforward physics and less peculiarity.

\subsection{Implementation of the three components of the exosphere}
We use the Rosetta/ALICE measurements of the hydrogen and oxygen exospheres to construct a spherically symmetric exosphere model for current study. \citet{feldman2011} provided tabulated density profiles for hydrogen, cold oxygen, and hot oxygen populations above 200 km. For each population, we first interpolate original profile of coarse ($\Delta d$ = 200 km) altitude resolution to that with a finer resolution ($\Delta d < 20 km$). Then we apply polynomial fitting to the logarithm of the density as a function of altitude within the altitude range shown in \citet{feldman2011}. Below 200 km the density is set to zero. Above the upper limit of the altitude range, we assume that the density decreases exponentially with the scale height determined by the highest 4 points of the interpolated density profiles. The formula is (all units are in SI unit) 
\begin{equation}
  \log_{10}(n(h)) = \sum^5_{i=0} a_i \cdot h^i
  \label{eq:nh}
\end{equation}

\begin{table}
\caption{Table of coefficients for the expression (\ref{eq:nh}) of number density as a function of height above Mars' surface. The height range for each set of coefficients are also given. }
\begin{tabular}{llll}
\hline
                  & $a_0$ & $a_1$ & $a_2$ \\ \hline
H, 200-35000 km   & 11.376328 & -5.9864716e-07 & 5.2802149e-14  \\
H, $>35000$ km    & 8.4164271 & -3.5031295e-08 & \\ 
Cold O, 200-1200 km & 16.272228 & -1.4563270e-05 & 2.6508755e-12 \\
Cold O, $>1200$ km  & 13.329357 & -8.9676294e-06 & \\
Hot O, 200-1400 km  & 11.456117 & -2.3702565e-06 & 4.0228317e-13 \\
Hot O, $>1400$ km   & 10.817852 & -1.3562674e-06 & \\ \hline 
                  & $a_3$ & $a_4$ & $a_5$ \\ \hline
H, 200-35000 km   & -2.6454801e-21 & 6.5495128e-29 & -6.2475058e-37 \\ \hline
\end{tabular}
\end{table}

\section{Conclusions}
We have presented the implementation of several model details for a 
parallel hybrid solver in the FLASH open source software framework.
First, an inclusion of permanent magnetic fields such that the divergence 
of the magnetic field is ensured to be zero in the discrete sense was 
presented. 
Secondly, a Monte Carlo algorithm for including ionospheric chemistry in the 
model is introduced, along with reaction times from the existing 
literature. 
Finally, an analytical fit for Mars hydrogen and oxygen exosphere, 
based on Rosetta observations, is presented.

\section*{Acknowledgements}
This research was conducted using resources provided by the Swedish National Infrastructure for Computing (SNIC) at the High Performance Computing Center North (HPC2N), Ume\aa\ University, Sweden.
The software used in this work was in part developed by the 
DOE NNSA-ASC OASCR Flash Center at the University of Chicago.

\end{document}